\documentstyle[sprocl,epsf,rotate,graphics]{article}

\def\NPB{{\em Nucl.}\ {\em Phys.} B}
\def\PLB{{\em Phys.}\ {\em Lett.}  B}
\def\PRL{{\em Phys.}\ {\em Rev.}\ {\em Lett.}}
\def\PRD{{\em Phys.}\ {\em Rev.} D}
\def\ZPC{{\em Z. Phys.} C}

\def\de{\partial}

\begin{document}

\title{BUBBLE WALL PROFILES IN SUPERSYMMETRIC MODELS$^a$}
\author{%
\hfill 
Peter John$^b$
\hfill\raisebox{21mm}[0mm][0mm]{\makebox[0mm][r]{HD-THEP-99-2}}%
}

\address{Institut f\"ur  Theoretische Physik, Philosophenweg 16,\\
69120 Heidelberg, Germany\\Email P.John@thphys.uni-heidelberg.de}

\maketitle\abstracts{We present solutions to the equations of motion for bubble wall profiles in 
the minimal and a non minimal supersymmetric extension of the Standard model. We discuss the 
method of the numerical approach and present results for the two models (MSSM and NMSSM). 
}
\footnotetext{${}^a$Presented at ``Strong and Electroweak Matter 98'', Copenhagen, 2.-5.12.1998}
\footnotetext{${}^b$Collaboration with S.J. Huber and M.G. Schmidt.}
\section{Introduction}
For the emergence of a  baryon asymmetry of the Universe the Sakharov
conditions necessarily demand deviation from thermodynamical
equilibrium. This condition is fulfilled in first order phase
transitions.  They take place via nucleation of bubbles separating the
symmetric from the broken phase. A first order phase transition might
occur at temperatures around the electroweak scale. It turned out that
in the Standard Model (SM) there is no phase transition at all for
Higgs masses larger than 72 GeV. Baryon number generation at the electroweak scale
therefore requires more complicated models with additional light
scalar fields such as the MSSM or NMSSM. In the MSSM there is a window for
electroweak baryogenesis and an upper bound for the Higgs mass of
about $m_H< 105$ GeV with a light stop\cite{bpj} of mass $m_{{\tilde t}_R}<m_{top}$ . 
In the NMSSM the bound on the Higgs mass is even
weaker \cite{frog}.

Having established the existence of a first order phase transition one
can start the actual calculation of the baryon asymmetry itself. There
are several mechanisms described in the literature.  All of them need
the knowledge of the profile of the bubble wall during the phase
transition. The kink ansatz in many situations is a good approximation
but it might be interesting to have a more refined description and to
determine which deviations occur in the presence of potentials
depending on two or more Higgs fields and eventually on CP violating
phases \cite{copi}.  Having the exact profile one can investigate the
dynamics of expanding bubbles and calculate the baryon asymmetry.

To determine the bubble wall profile beyond a simple ansatz we have
to  solve the equations of motion numerically. In the case of more than
one scalar field this is a highly nontrivial task since simple methods
like  over\-shoot\-ing-under\-shoot\-ing fail.  So one has to use methods
which, beginning with an ansatz, converge to the actual solution. They
are sometimes called ``relaxation methods''. 

We first have to find the equations of motion.  In field theory they
can be derived via Euler Lagrange equations from the Lagrangian
density which has the general form
$ {\cal L} = (D_\mu \Phi_i)^+(D^\mu\Phi_i) + V(\Phi_i,T)$
for several Higgs fields $\Phi_i$ (plus, eventually, CP violating
phases). Here $D_\mu$ is a covariant derivative and $V$ denotes the
effective potential. We show results of investigations of the MSSM and
the NMSSM. 
\subsection{Critical Bubble }
At the nucleation temperature the free energy becomes positive and the bubbles start
to grow until they fill up the entire space. The profile of the radial
symmetric bubbles along the radius $r$ is determined by the equations
of motion for the critical bubble (``bounce''):
\begin{equation}
0  =\frac{\de^2\phi_i}{\de r^2}+\frac{2}{r}\frac{\de\phi_i}{\de r}  - \frac{\de V}{\de \phi_i}\quad i=1\ldots N
\end{equation}
with boundary conditions  $\frac{\de\phi_i}{\de r}=0|_{r=\infty}$ and $\quad\phi_i=0|_{r=\infty}$, where  $N=2, 3$ for the MSSM and NMSSM respectively.
\subsection{Stationary Bubble or Domain Walls}
Constraining to a stationary wall with velocity $v_w$ at a late time $t$
where the wall is already almost flat we are left with only one
spatial dimension $x=z-v_wt$ perpendicular to the wall and have the equations
\begin{equation}
0  =\frac{\de^2\phi_i}{\de x^2} - \frac{\de V}{\de \phi_i} :=E_i(x)\label{Ei}
\end{equation}
with two boundary conditions, e.~g. $\frac{\de\phi_i}{\de x}=0|_{x=\infty}$ and $\quad \phi_i=0|_{x=-\infty}$. We will discuss now the method and show solutions calculated with it.
\section{Solutions}
\subsection{The Method}
Since the  over\-shoot\-ing-under\-shoot\-ing method fails we have to
devise another method.  We here minimize the
functional of squared equations of motion.  Then, solving the eqs. of motion
means finding field configurations for which the functional
\begin{equation}
{\cal F}=\int_{-\infty}^{+\infty}dx \left(E_1^2(x)+E_2^2(x)+\ldots+ E_N^2(x)\right)\nonumber
\end{equation}
is zero, which is achieved by minimizing ${\cal F}$.  This method
has already been used\cite{span,moore} for the critical bubble of the
MSSM.

Applying the minimization method we have to solve a boundary
value problem. Thus we have to make an ansatz for every function for
which we want to find the time development which fulfills the boundary
conditions. 
\subsection{The Algorithm}
The numerical algorithm works in two steps:
\begin{enumerate}
\item Find an ansatz as close as possible to the exact solution by low dimensional minimization of ${\cal F}$ with special ansatz configurations.
\item High dimensional minimization of  ${\cal F}$ by discretizing every field function on a grid and minimization with respect to the function values.
\end{enumerate}
In the MSSM the kink ansatz is quite appropriate to get a good
starting configuration for the minimization procedure of step 1. In
the NMSSM the improved kink ansatz is not that good but nevertheless
still appropriate.  With only a few parameters $L_i$ and ${\hat x}_i$
this a low dimensional minimization procedure.  In the MSSM we only
have two ($L_1$, $L_2$), the offset ${\hat x}$ is negligible, in the NMSSM five parameters ($L_{1/2/3}$,
${\hat x}_{1/2}$, ${\hat x}=0$). We use as ansatz configurations:
\begin{equation}
\phi_i^{kink}=\frac{v_i}{2}\left(1+\tanh(\frac{x}{L_i}+{\hat x}_i)\right),\ i=1\ldots N
\end{equation}
With $N$ fields on a grid with $M$ space points the second step is a
$N\times M$ dimensional minimization, which must be performed with
fast converging methods. The results here are obtained by $N=3$
fields, $M\sim 60-100$ grid points and Powell's quadratically
converging method.  The derivatives of the differential equations are
discretized with three and four point formulae, the integrals of
${\cal F}$ are performed with an extended Simpson rule. The field
configurations are interpolated by splines \cite{pj,numrec}. For more
details on the algorithm, see also \cite{pj}. The worst problem doing
numerics is the existence of spurious minima. Besides real solutions
to the equations there are fake minima due to the numerical
representation and solutions due to the fact, that $\delta E^2=0$ is
fulfilled not only for $E=0$ but also for $\delta E= 0$. One can
perform checks to rate the minima found (see \cite{pj,AK}).
\subsection{Applications}
We will show now applications of the described method. First, in figure 1
 we present solutions of the bubble wall profile. The deviation
from the straight line is small but nevertheless responsible for the
actual amount of baryon asymmetry. Figure 2 shows the same for
the three field case of the NMSSM. There we have a considerable
deviation from the straight line and, additionally, a stronger
deviation from the the extended kink ansatz. This demonstrates also the
importance of a general solution method. Figure 2 shows also the path
of the mechanical analogue of a rolling marble along the ridge of the
potential.\\
\begin{figure}
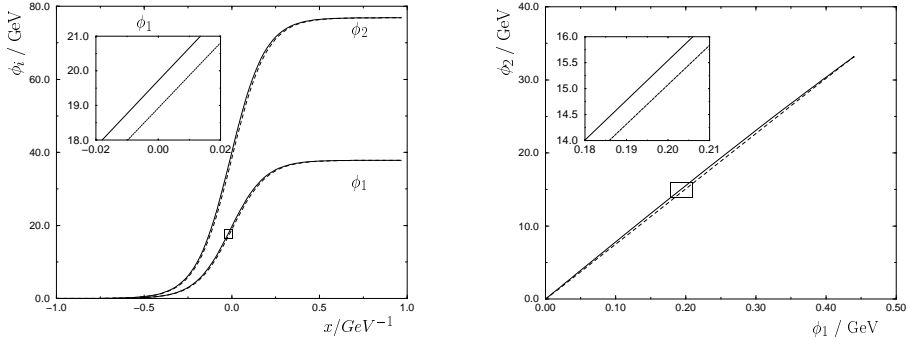

\vspace*{5cm}
\begin{picture}(0,0)
\put(10,0){\epsfysize=6cm\epsffile{fig1ges.eps}}
\put(195,0){\epsfysize=6cm\epsffile{fig3ges.eps}}
\end{picture}
\vspace*{-1.5cm}
\caption{Solutions for the MSSM. Left: Solution (solid) configuration and $tanh$-ansatz (dashed). Only for unphysically small values of $m_A\approx{\cal O}$(10 GeV) the deviation from the straigth line (dashed) becomes appreciable. Wall widths are around $L\approx 10/T_c-30/T_c$ GeV${}^{-1}$.}
\end{figure}
\begin{figure}
\vspace*{0cm}
\begin{picture}(0,0)
\put(10,-200){\epsfysize=6cm\epsffile{fig1nges.eps}}
\put(100,-50){$\phi_1$}
\put(100,-80){$\phi_2$}
\put(100,-117){$\phi_3$}
\put(170,-150){\epsfysize=6cm\rotate[r]{\epsffile{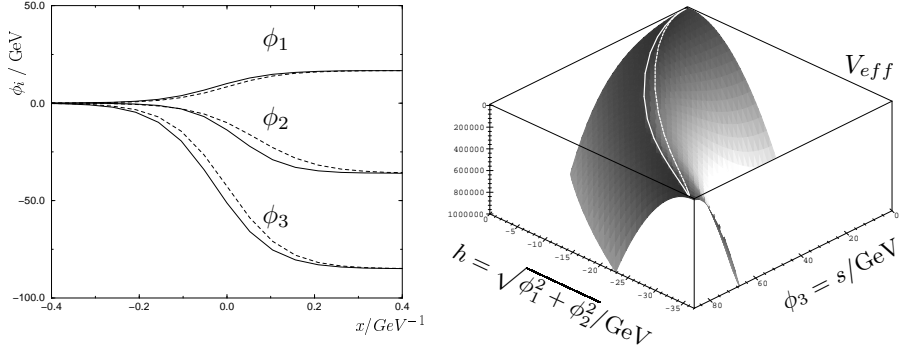}}}
\put(320,-60){$V_{eff}$}
\put(170,-130){\rotatebox{335}{\small$h=\sqrt{\phi_1^2+\phi_2^2}$/GeV}}
\put(295,-150){\rotatebox{27}{\small $\phi_3=s$/GeV}}
\end{picture}
\label{fig2}
\vspace*{6cm}
\caption{Left: Solutions (solid) for the NMSSM compared to $tanh$-ansatz  with offsets (dashed). The deviation from the ansatz is significant. Wall widths are around $L\approx 15/T_c$ GeV${}^{-1}$. Right: Tunneling configuration on the negative potential; path of ``marble'' analogue. $\phi_3=s$ is the singlet field of the NMSSM.}
\end{figure}
These results indicate the general behaviour of solutions in theories
with more than one Higgs field. Now questions of metastability of
wrong minima can be investigated with higher accuracy and better
reliability.  This is one step for a more precise calculation of the
actual amount of the baryon asymmetry of the universe.
\section*{Acknowledgments}
I am grateful for a lot of discussions with M. Schmidt, S. Huber, M. Laine, D. B\"odeker, G. Moore, M. Seco and C. Wetterich. Partially supported by TMR network, EU contract no. ERBFMRXCT97-0122.


\section*{References}
\begin{thebibliography}{99}
\bibitem{bpj} D.~B\"odeker,\  P.~John,\  M.~Laine,\  M.G.\ Schmidt,\  \NPB497(1997)387; M.~Carena,\  M.~Quir\'os,\  C.E.M.~Wagner,\  \NPB524(1998)3; M.~Losada,\  \NPB537(1999)3; M.~Laine,\  K.~Rummukainen,\  {\em Phys.}\ {\em Rev.}\ {\em Lett.}80(1998)5259,  \NPB535(1998)423; J.M.\ Cline hep-ph/9810267; J.M.\ Cline,  G.D.\ Moore,  \PRL81(1998)3315
\bibitem{frog} A.D.~Davies,\ C.D.~Froggatt,\ R.G.~Moorhouse, \PLB372(1996)88;  S.J.~Huber,\  M.G.~Schmidt,\  hep-ph/9809506, to appear in {{\em Eur.\ J.} C}; contribution by S.J.~Huber, these proceedings, S.J.~Huber, P.~John, M.G.~Schmidt, work in progress.
\bibitem{copi}D.~Comelli,  M.~Pietroni,  \PLB306(1993)67; A.~Hammerschmitt,\ J.~Kripfganz,\ M.G.~Schmidt,\ \ZPC64(1994)105; J.~Cline,  K.~Kainulainen,  A.P.~Visher,  \PRD54(1996)2451; K.~Funakubo,\ hep-ph/9809517;  M.~Laine, K.~Rummukainen, hep-ph/9810369
\bibitem{span} J.~M.~Moreno,\ M.~Quir\'os,\ M.~Seco, \NPB526(1998)489
\bibitem{pj} P. John, hep-ph/9810499
\bibitem{numrec}  W.~Press et al.,  {\em Numerical recipes}, Cambridge University Press 1988; M.~Abramowitz,\ I.A.~Stegun,\ {\em Handbook Of Mathematical Functions With Formulas, Graphs and Mathematical Tables}, John Wiley \& Sons, 1964,1972
\bibitem{AK} A.~Kusenko, \PLB358(1995)47,51
\bibitem{moore} Another method is described in the appendix of hep-ph/9810261, which  has been pointed out to me by G.\ Moore.
\end{thebibliography}
\end{document}